\documentclass[journal]{IEEEtran}

\usepackage{amsmath,amssymb,amsfonts}
\usepackage{algorithm}
\usepackage{algorithmic}
\usepackage{array}
\usepackage{booktabs}
\usepackage{flushend}
\usepackage{graphicx}
\usepackage{cite}
\usepackage{textcomp}
\usepackage{url}

\newcommand{\jj}{\mathrm{j}}
\newcommand{\T}{\mathsf{T}}

\newcommand{\bc}{\mathrm{bc}}
\newcommand{\net}{\mathrm{net}}
\newcommand{\ext}{\mathrm{ext}}
\newcommand{\raw}{\mathrm{raw}}
\newcommand{\diag}{\operatorname{diag}}
\newcommand{\rank}{\operatorname{rank}}
\newcommand{\vect}[1]{\boldsymbol{#1}}
\newcommand{\mat}[1]{\mathbf{#1}}

\begin{document}

\title{Sparse Generalized-Admittance AC Power Flow for Fast Contingency Analysis and Remedial-Action Assessment}

\author{
    \IEEEauthorblockN{Pierre Artoisenet and No\'emie Verstraete}\\
    \IEEEauthorblockA{N-SIDE\\
    1348 Louvain-la-Neuve, Belgium\\
    Email: \{pierre.artoisenet, noemie.verstraete\}@n-side.com}\\
    September 2026
}


\maketitle

\begin{abstract}
Repeated AC power-flow calculations are common tasks in transmission system operations,
in particular for  contingency analysis and remedial-action assessment.
 For large networks, their cumulative runtime can become a computational bottleneck
in operational processes.

This paper develops a sparse
reformulation of the generalized bus-admittance power-flow method known as PFPD. Loads and
generators are represented by fixed shunt admittance terms on the diagonal of the admittance
matrix, while corrective nodal currents are iteratively computed to match the
constant-power and regulated-voltage constraints. In contrast to
the original formulation, the slack bus is treated as a fixed-voltage boundary and removed
from the admittance block decomposition. The resulting iteration reuses sparse LU factorizations of
the initial generalized non-slack matrix and its PQ block, without explicitly forming dense
inverse matrices. The same formulation admits two initialization modes: an unsolved base
case using the same flat-start shunt heuristic as in the primary PFPD method, and repeated
post-action calculations initialized from a solved operating point. Localized topology and parameter changes are expressed as
low-rank admittance updates and the resulting admittance matrices are inverted by means of the Woodbury identity. An optional
local-network formulation can provide an warm start before full-network refinement.
N-1 experiments on the 1354-bus and 9241-bus PEGASE systems and the 6717-bus
synthetic Texas system yield average per-contingency speedups between 14.4 and
18.0 relative to pandapower. On transformer tap-position actions, the method is
2.8--2.9 times faster than pandapower and 7.2--29.9 times faster than PowSyBl OpenLoadFlow.
\end{abstract}


\section{Introduction}

\IEEEPARstart{A}{C} power flow is a basic computational task in power-system planning,
operation, security analysis, and optimization. Given the network model and the external
quantities at slack, PV, and PQ buses, the problem determines the steady-state bus-voltage
phasors and the remaining active and reactive injections. Operational applications rarely
require a single calculation: N-1 screening, remedial-action optimization, voltage-control
studies, and uncertainty analysis may require hundreds or thousands of closely related
solutions on the same large network \cite{fliscounakis2013contingency}.

Early current- or admittance-based successive-substitution methods have computationally cheap
iterations but may converge slowly. Newton--Raphson (N--R) power flow instead solves
a linearized nonlinear mismatch problem and is praised for its local quadratic
convergence \cite{tinney1967newton,grainger1994power}. Its main recurring costs are the
construction of the Jacobian matrix and its inversion at each iteration. The fast-decoupled load flow (FDLF)
freezes and decouples approximations of the active-power/angle and reactive-power/magnitude
Jacobian blocks, reducing the work per iteration under the usual transmission-grid
assumptions \cite{stott1974fast}. Mature open-source implementations include
MATPOWER, pandapower and OpenLoadFlow \cite{zimmerman2011matpower,thurner2018pandapower, powsyblopenloadflow}.

Benato's Power Flow of the University of Padova (PFPD) follows a different
approach \cite{benato2001complex,benato2022basic}. It embeds
constant-power loads and generators as approximated shunt admittance terms in a generalized bus-admittance
matrix. This generalized admittance matrix is held fixed, and nonlinear power constraints are restored exactly by
iteratively injecting corrective nodal currents. The 2022 formulation additionally
represents the slack generator by a Norton equivalent with a very large shunt admittance and a very large nodal current.
It exhibits robust convergence in the reported tests, but the proposed construction contains a Schur complement followed by another matrix inversion.
The matrix used to resolve the currents at the PV buses is generally dense even though the underlying bus-admittance matrix is sparse.
This limits the direct use of sparse triangular solves and becomes increasingly costly as the grid and the number of PV buses grow (typically grids with more than 10k nodes).

This paper reformulates PFPD for sparse computation and for repeated changes around a common operating point. Its contributions are as follows.

\begin{enumerate}
\item The slack bus is removed from the admittance matrix decomposition and retained as a
fixed-voltage boundary. The non-slack voltages are obtained by inverting a sparse linear system
built upon a generalized admittance matrix.
\item A block-inverse identity is used to replace the dense voltage to current PV operator in the PFPD
approach by an alternative formulation avoiding any explicit dense matrix.
Each nonlinear iteration uses triangular solves based on reusable sparse LU factors of the non-slack admittance matrix and its PQ block,
together with sparse matrix products.
\item The method is presented both as a stand-alone base-case solver, initialized
with the PFPD flat-start shunt heuristic, and as an accelerated post-action solver initialized
from a solved base case.
\item Branch disconnections, transformer or phase-shifting-transformer tap changes,
substation reconfigurations, and diagonal shunt changes are expressed as localized,
low-rank admittance matrix changes. The Woodbury identity is used to compute the post-action inverse operators
without refactorizing the full matrices.
\item A boundary-voltage subnetwork formulation is described as an optional local warm start
before full-network refinement; its performance is left for future evaluation.
\item Experiments on three networks report execution time relative to pandapower and
PowSyBl OpenLoadFlow, together with voltage agreement relative to pandapower.
\end{enumerate}

The paper focuses on balanced positive-sequence power flow.
The test cases presented in this paper assume a load-flow mode without reactive-power limits,
while an efficient handling of reactive power limits at PV nodes will be presented in another work.
Section~II defines the generalized model.
Section~III derives the sparse iteration.
Sections~IV--VI discuss implementation, actions, and locality.
Section~VII presents the experimental protocol and results, and Section~VIII concludes.

\section{Generalized-Admittance Power-Flow Model}
\label{sec:model}

\subsection{Conventions and Bus Partition}

All quantities are expressed in per unit system. Bold lowercase and uppercase symbols denote complex
vectors and matrices, respectively; $(\cdot)^*$ denotes complex conjugation.
Net external complex power $s_k^{\mathrm{ext}}=p_k^{\mathrm{ext}}+\jj q_k^{\mathrm{ext}}$
is positive for consumption and negative for generation.
Using these conventions, the power flow equations read
\begin{equation}
\vect{u} \odot  \left( \mat{Y}^{\net} \vect{u} \right)^* = - (p^{\mathrm{ext}}+\jj q^{\mathrm{ext}})
\label{eq:pf_equations_with_classic_Y}
\end{equation}
with $ \mat{Y}^{\net}$ the classical bus admittance matrix.

Let $s$ denote the single slack bus,
$\mathcal V$ the PV buses, $\mathcal Q$ the PQ buses,
and $\mathcal L=\mathcal V\cup\mathcal Q$ all non-slack buses.
Without loss of generality, the following ordering is considered: first the slack bus,
then the PV buses and finally the PQ buses.
The physical network-admittance matrix is hence partitioned as
\begin{equation}
\mat{Y}^{\net}=
\begin{bmatrix}
Y_{ss}^{\net} & \mat{Y}_{s\ell}^{\net}\\
\mat{Y}_{\ell s}^{\net} & \mat{Y}_{\ell\ell}^{\net}
\end{bmatrix},
\qquad
\mat{Y}_{\ell\ell}^{\net}=
\begin{bmatrix}
\mat{Y}_{vv}^{\net} & \mat{Y}_{vq}^{\net}\\
\mat{Y}_{qv}^{\net} & \mat{Y}_{qq}^{\net}
\end{bmatrix}.
\label{eq:partition}
\end{equation}
where the letters used for the subscripts $s, l, v, q$ indicate the nature of the associated buses.
In the previous Equation, we used the subscript $\net$ to refer to the
classical bus admittance matrix, whereas the same notations without the subscript $\net$
is used for the generalized admittance matrix, introduced in the next subsection.

\subsection{The generalized admittance matrix}

The generalized admittance matrix $\mat{Y}$ differs from the classical admittance matrix $\mat{Y}^{\net}$ by an additive vector
$\vect{y}^{\ext}$ of shunt admittances contributing on the diagonal of the admittance matrix associated with non-slack buses:
\begin{equation}
\mat{Y}_{\ell\ell}=\mat{Y}_{\ell\ell}^{\net}
+\diag(\vect{y}_{\ell}^{\ext}).
\label{eq:generalized_y}
\end{equation}
Inserting this last expression into (\ref{eq:pf_equations_with_classic_Y}),
the power flow equations for non-slack buses can be expressed with the generalized admittance matrix:
\begin{equation}
\vect{u}_\ell \odot  \vect{i}_\ell^* - |\vect{u}_\ell|^2 \odot (\vect{y}_\ell^{\ext })^* = - (\vect{p}^{\mathrm{ext}}+\jj \vect{q}^{\mathrm{ext}})_\ell
\label{eq:pf_equations_with_generalized_Y}
\end{equation}
where $\vect{i} = \mat{Y} \vect{u}$ denotes the nodal currents resulting from the generalized admittance matrix.
The vector of shunt admittances is designed to approximate the external powers on the right-hand side of (\ref{eq:pf_equations_with_generalized_Y}).
In the ideal case
\begin{equation}
\vect{y}_{\ell}^{\ext, \textrm{ideal}} = \frac{(\vect{p}^{\mathrm{ext}} - \jj \vect{q}^{\mathrm{ext}})_\ell}{|\vect{u}_\ell|^2}
\label{eq:y_ideal}
\end{equation}
where division of vectors is applied element-wise, and the power flow equations with generalized admittance
matrix are satisfied with zero corrective current $\vect{i}_{\ell}$ at every non-slack bus, as it can be seen from (\ref{eq:pf_equations_with_generalized_Y}).

In practice however, the nodal voltage magnitudes at PQ buses and reactive power injections at the PV buses are not known before the power-flow equations are solved. Consequently, approximate shunt admittances are computed using reference voltage magnitudes $|u^{\mathrm{ref}}|$ and initial reactive power estimates $q_0$:
\begin{equation}
\vect{y}_{\ell}^{\ext} = \frac{(\vect{p}^{\mathrm{ext}} - \jj \vect{q_0})_\ell}{|\vect{u}^{\mathrm{ref}}_{ \ell}|^2}
\label{eq:y_ext_general}
\end{equation}

Unlike an iterative impedance-load model,  $\vect{y}^{\ext}$ remains fixed
throughout one entire power flow solve cycle. In this way, the generalized admittance
matrix stays constant, which lowers the computation cost per iteration.
The error behind approximate shunt admittances is compensated by the adaptation of the
corrective-current vector $\vect{i}_{\ell}$:
\begin{equation}
\mat{Y}_{\ell\ell}\vect{u}_{\ell}
+\mat{Y}_{\ell s}u_s=\vect{i}_{\ell}.
\label{eq:corrected_kcl}
\end{equation}
which is updated at each iteration.
At convergence, the shunt admittance and the corrected
current matches exactly the external power at each node, implying that equation (\ref{eq:pf_equations_with_generalized_Y}) is satisfied.

\subsection{Shunt Admittances at PQ and PV buses}

For an unsolved base case, PFPD's initialization is retained
without the large slack shunt introduced in \cite{benato2022basic}.
At PQ buses, the shunt admittances are set to (\ref{eq:y_ext_general}) with $q_0 = q^{\mathrm{ext}}$
and reference voltage magnitudes $|u_q^{\mathrm{ref}}|$ that can differ from 1 p.u.
\begin{equation}
\vect{y}_q^{\ext}=\frac{\vect{p}_q^{\mathrm{ext}}-\jj \vect{q}_q^{\mathrm{ext}}}{|\vect{u}_q^{\mathrm{ref}}|^2}.
\label{eq:pq_initial_shunt}
\end{equation}
At the PV buses, with voltage magnitude targets
 $V_v^{\mathrm{set}}$, and an initial reactive-power estimate $q_0 = q_v^{0}$,
 the shunt admittance is\footnote{As a reminder, power is positive for consumption and negative for generation.}
\begin{equation}
\vect{y}_v^{\ext}=
\frac{\vect{p}_v^{\mathrm{ext}} - \jj \vect{q}_v^{0}}{(\vect{V}_v^{\mathrm{set}})^2}.
\label{eq:pv_initial_shunt}
\end{equation}
As in~\cite{benato2022basic}, the estimates $q_v^{0}$ are obtained from an auxiliary flat-start calculation:
retain the imaginary part of $\mat{Y}^{\net}$, include the PQ
shunts of~\eqref{eq:pq_initial_shunt}, impose PV magnitudes with zero angles,
compute the resulting generator currents, and infer their reactive powers.
Benato reports a broad region of attraction around this estimate \cite{benato2022basic}.

For repeated studies around a solved base-case operating point, the proposed approach in this
paper is to consider the ideal base case shunt admittances defined in (\ref{eq:y_ideal})
with the voltage and reactive power profiles associated with the base case power flow solution.
\begin{equation}
\vect{y}^{\ext,\bc}_\ell=\frac{\vect{p}_{\ell}^{\mathrm{ext}}-\jj \vect{q}_\ell^{\mathrm{bc}}}{|\vect{u}_\ell^{\mathrm{bc}}|^2}.
\label{eq:bc_shunt}
\end{equation}
Such a choice makes zero corrective current reproduce the solved base case exactly.
It provides a useful warm start for the assessment of remedial actions
or contingencies.

\subsection{Slack Elimination}

The slack phasor $u_s=V_s^{\mathrm{set}}e^{\jj\theta_s^{\mathrm{set}}}$ is fixed. Solving~\eqref{eq:corrected_kcl} gives
\begin{align}
\vect{u}_{\ell}
&=\vect{u}_{\ell}^{0}+\mat{Y}_{\ell\ell}^{-1}\vect{i}_{\ell},
\label{eq:voltage_from_current}\\
\vect{u}_{\ell}^{0}
&=-\mat{Y}_{\ell\ell}^{-1}\mat{Y}_{\ell s}u_s.
\label{eq:zero_current_voltage}
\end{align}
The superscript $0$ is used above to denote the voltage profile resulting from the fixed generalized matrix without any corrective currents, not necessarily a flat voltage profile.
Numerically, the products with $\mat{Y}_{\ell\ell}^{-1}$ are expressed as sparse linear solves; the inverse is never assembled.
Removing the slack bus also removes the arbitrarily large Norton admittance of the
all-inclusive PFPD model and avoids the associated scaling choice for the large current at the slack bus.

\section{Sparse Iterative Solution}
\label{sec:iteration}

Define the solve operators
\begin{equation}
\mathcal S_{\ell}(\vect{b})=\mat{Y}_{\ell\ell}^{-1}\vect{b},
\qquad
\mathcal S_q(\vect{b})=\mat{Y}_{qq}^{-1}\vect{b},
\label{eq:solve_operators}
\end{equation}
implemented by means of reusable LU factors. One iteration maps the previous corrective currents to updated currents and voltages as follows.

First, the non-slack voltages are updated as follows:
\begin{equation}
\vect{u}_{\ell}^{(k)}=\vect{u}_{\ell}^{0}
+\mathcal S_{\ell}\!\left(\vect{i}_{\ell}^{(k-1)}\right).
\label{eq:iteration_voltage}
\end{equation}
The PV magnitudes are projected onto their target set points while the angles are unchanged:
\begin{equation}
\vect{u}_v^{(k)}\leftarrow
\vect{V}_v^{\mathrm{set}}\odot
\frac{\vect{u}_v^{(k)}}{|\vect{u}_v^{(k)}|}.
\label{eq:pv_projection}
\end{equation}
As a reminder, all divisions and $\odot$ products between vectors are applied element-wise.

To identify the PV currents that produce the projected voltage,
 the voltage contribution of the previous PQ currents is isolated:
\begin{align}
\Delta \vect{u}^{(q,k)}
&=\mathcal S_{\ell}
\begin{bmatrix}\vect{0}\\\vect{i}_q^{(k-1)}\end{bmatrix},
\label{eq:pq_contribution}\\
 \widetilde{\vect{u}}_v^{(k)}
&=\vect{u}_v^{(k)}-\vect{u}_v^{0}- \Delta \vect{u}_v^{(q,k)}.
\label{eq:pv_residual_voltage}
\end{align}
If $\mat{Z}=\mat{Y}_{\ell\ell}^{-1}$, then $\widetilde{\vect{u}}_v=\mat{Z}_{vv}\vect{i}_v$.
The inverse of this generally dense subblock is never constructed. Instead, block inversion allows to express
the inverse in terms of sparse admittance subblocks and inverse of PQ block:
\begin{equation}
\mat{Z}_{vv}^{-1}=\mat{Y}_{vv}
-\mat{Y}_{vq}\mat{Y}_{qq}^{-1}\mat{Y}_{qv},
\label{eq:block_inverse_identity}
\end{equation}
and hence
\begin{equation}
\vect{i}_v^{\raw,(k)}=
\mat{Y}_{vv}\widetilde{\vect{u}}_v^{(k)}
-\mat{Y}_{vq}\mathcal S_q
\!\left(\mat{Y}_{qv}\widetilde{\vect{u}}_v^{(k)}\right).
\label{eq:pv_raw_current}
\end{equation}

Combining these updated currents at PV buses with the previous currents at PQ buses gives
\begin{equation}\vect{i}_\ell^{\raw,(k)}=\begin{bmatrix}\vect{i}_v^{\raw,(k)}\\ \vect{i}_q^{(k-1)}\end{bmatrix}.
\label{eq:raw_full_current}
\end{equation}
The PQ voltages are then recomputed as
\begin{equation}\vect{u}_q^{(k)}=\vect{u}_q^0+\left[\mathcal S_\ell\!\left(\vect{i}_\ell^{\raw,(k)}\right)\right]_q,
\label{eq:pq_voltage_recompute}
\end{equation}
while the projected PV voltages from~\eqref{eq:pv_projection} are retained.
At this stage, all voltages at non-slack buses are updated, and the algorithm
proceeds with finalizing the corrective currents.

The corrective currents are adjusted to enforce the constant-power constraint in Equation~\eqref{eq:pf_equations_with_generalized_Y}.
For PQ buses, Equation~\eqref{eq:pq_initial_shunt} implies that the right hand side of \eqref{eq:pf_equations_with_generalized_Y} can be substituted by
$ - \left(\vect{y}_q^{\ext}\right)^* \odot |\vect{u}_q^{\mathrm{ref}}|^2$, so that the corrective current simplifies into
\begin{equation}
\vect{i}_q^{(k)}=
\vect{y}_q^{\ext}\odot
\frac{|\vect{u}_q^{(k)}|^2-|\vect{u}_q^{\mathrm{ref}}|^2}
{(\vect{u}_q^{(k)})^*}.
\label{eq:pq_corrective_current}
\end{equation}

For PV nodes, taking the complex conjugate of Equation~\eqref{eq:pf_equations_with_generalized_Y} and decomposing into real and imaginary parts yields
\begin{align}
\Re \left( \vect{u}_v^{(k)*} \odot  \vect{i}_v^{(k)} \right) &= -\vect{p}^{\mathrm{ext}}_v + |\vect{u}_v^{(k)}|^2 \odot \Re \left(\vect{y}_v^{\ext}\right) \label{eq:pv_corrective_current_1}
 \\
\Im \left( \vect{u}_v^{(k)*} \odot  \vect{i}_v^{\raw,(k)} \right) &= \vect{q}_v + |\vect{u}_v^{(k)}|^2 \odot \Im \left(\vect{y}_v^{\ext }\right)
\label{eq:pv_corrective_current_2}
\end{align}
The known voltage magnitude target in the expression of the shunt admittance term for PV node in \eqref{eq:pv_initial_shunt}
implies that the left-hand side of Equation \eqref{eq:pv_corrective_current_1} is zero, which is enforced in the updated corrective current
$\vect{i}_v^{(k)}$.  Equation~\eqref{eq:pv_corrective_current_2} instead is expressed in terms of the raw current $\vect{i}_v^{\raw,(k)}$
and is used to extract the reactive power $\vect{q}_v$ at the PV nodes.
The final expression for $\vect{i}_v^{(k)}$ ensures that only the external reactive power is adjusted,
i.e. the corrective current  must carry only quadrature power.
Projecting the raw current onto that component gives
\begin{equation}
\vect{i}_v^{(k)}=\jj\,
\frac{\Im\!\left\{\vect{u}_v^{(k)*}\odot
\vect{i}_v^{\raw,(k)}\right\}}
{(\vect{u}_v^{(k)})^*}.
\label{eq:pv_quadrature_current}
\end{equation}
The complex power associated with~\eqref{eq:pv_quadrature_current} is purely imaginary,
so the active generation constraint in Equation~\eqref{eq:pv_corrective_current_1} is satisfied.

Finally, the power gap at each iteration can be computed as the mismatch between the
left-hand side 	and right-hand side of Equation~\eqref{eq:pf_equations_with_generalized_Y}
when considering the raw corrective currents $\vect{i}_l^{\raw,(k)}$, so before applying the power constraint update:
\begin{equation}
\mathrm{gap}^{(k)} =
\vect{u}_{\ell}^{(k)} \odot  \vect{i}_l^{\raw,(k)*} - |\vect{u}_{\ell}^{(k)}|^2 \odot (\vect{y}_\ell^{\ext })^*  + (\vect{p}^{\mathrm{ext}}+\jj \vect{q}^{\mathrm{ext}})_\ell
\end{equation}
where the reactive power for PV buses is set to satisfy equation~\eqref{eq:pv_corrective_current_2}.

Algorithm~\ref{alg:sparse_pfpd} summarizes the stand-alone base-case path. The post-action path uses the
same loop after replacing the base matrices and zero-correction voltage by their action-updated operators
in Section~\ref{sec:actions}.

\begin{algorithm}[t]
\caption{Sparse generalized-admittance AC power flow}
\label{alg:sparse_pfpd}
\begin{algorithmic}[1]
\REQUIRE $\mat{Y}^{\net}$, bus partition, specified powers, voltage set points, $u_s$, tolerance $\epsilon$, maximum iteration count $k_{\max}$
\STATE Build $\vect{y}^{\ext}$ using~\eqref{eq:pq_initial_shunt}--\eqref{eq:pv_initial_shunt}
\STATE Factorize sparse $\mat{Y}_{\ell\ell}$ and $\mat{Y}_{qq}$
\STATE Compute $\vect{u}_{\ell}^{0}$ from~\eqref{eq:zero_current_voltage}; set $\vect{i}_{\ell}^{(0)}=\vect{0}$
\FOR{$k=1,\ldots,k_{\max}$}
\STATE Compute $\vect{u}_{\ell}^{(k)}$ using~\eqref{eq:iteration_voltage}
\STATE Project $\vect{u}_v^{(k)}$ using~\eqref{eq:pv_projection}
\STATE Compute $\vect{i}_v^{\raw,(k)}$ using~\eqref{eq:pq_contribution}--\eqref{eq:pv_raw_current}
\STATE Recompute $\vect{u}_q^{(k)}$ with the updated PV current
\STATE Update $\vect{i}_q^{(k)}$ using~\eqref{eq:pq_corrective_current}
\STATE Update $\vect{i}_v^{(k)}$ using~\eqref{eq:pv_quadrature_current}
\IF{$\|\mathrm{gap}^{(k)}\|_{\infty}\leq\epsilon$}
\STATE \textbf{break}
\ENDIF
\ENDFOR
\RETURN bus-voltage phasors, currents, powers, convergence status
\end{algorithmic}
\end{algorithm}

The implementation monitors the infinity norm of the nodal complex mismatch gap.
The reported experiments use a target of $10^{-2}$ MVA, similarly to the default tolerance value in the OpenLoadFlow provider of PowSyBl.

\section{Numerical Implementation}
\label{sec:numerics}

The prototype stores the matrices in compressed sparse-column form and uses
SuperLU through SciPy for sparse partial-pivoting
LU factorization \cite{li2005superlu,virtanen2020scipy}. The two large factorizations are
\begin{equation}
\mat{P}_{\ell}\mat{Y}_{\ell\ell}\mat{Q}_{\ell}
=\mat{L}_{\ell}\mat{U}_{\ell},
\qquad
\mat{P}_{q}\mat{Y}_{qq}\mat{Q}_{q}
=\mat{L}_{q}\mat{U}_{q},
\label{eq:lu}
\end{equation}
where permutations limit fill and support stable pivoting. The factors are independent
of the nonlinear iteration count, hence these two factorization must be derived only
once independently of the number of iterations.
Two right-hand sides---the voltage effects of all corrective currents
and of the PQ currents alone---are solved together where possible.

Table~\ref{tab:comparison} contrasts the computational structures
in this approach and in the native PFPD paper. In the later approach,
the matrix\footnote{In the PFPD paper, the index $G$ is used for PV nodes, the index $L$ is used for
PQ nodes.}
\begin{equation}
\mat{Y}_{G,\mathrm{eq}}=\mat{Y}_{GG}
-\mat{Y}_{GL}\mat{Y}_{LL}^{-1}\mat{Y}_{LG}
\label{eq:benato_schur}
\end{equation}
is calculated and then inverted:
 $\mat{Z}_{G,\mathrm{eq}}=\mat{Y}_{G,\mathrm{eq}}^{-1}$.
Even if the solve with $\mat{Y}_{LL}$ is sparse, $\mat{Y}_{G,\mathrm{eq}}$ and
its inverse are normally dense. Equation~\eqref{eq:pv_raw_current} applies the
corresponding operator to a vector without forming either dense matrix.
The relevant cost is therefore primarily governed by sparse-factor fill,
 not by the square or cube of the bus count.

\begin{table}[t]
\caption{Computational Structure of the Two PFPD Formulations}
\label{tab:comparison}
\centering
\footnotesize
\begin{tabular}{p{0.22\columnwidth}p{0.29\columnwidth}p{0.29\columnwidth}}
\toprule
Property & Original PFPD & Proposed formulation\\
\midrule
Slack model & Large Norton shunt inside $\mat{Y}$ & Fixed-voltage boundary outside $\mat{Y}_{\ell\ell}$\\
Main operators & $\mat{Y}_{LL}^{-1}$ and $\mat{Y}_{G,\mathrm{eq}}^{-1}$ & Sparse solves with $\mat{Y}_{\ell\ell}$ and $\mat{Y}_{qq}$\\
Dense object & PV operator and its inverse & Only small action-update matrices\\
Reusable work & Fixed inverses within one solve & Sparse LU factors across iterations and actions\\
Action update & Not developed for repeated actions & Low-rank Woodbury update\\
\bottomrule
\end{tabular}
\end{table}

\section{Remedial Actions and Contingencies}
\label{sec:actions}

\subsection{Localized Admittance Changes}

Let an action modify the generalized matrix by
\begin{equation}
\mat{Y}'=\mat{Y}+\Delta\mat{Y}
=\mat{Y}+\mat{U}\mat{C}\mat{V},
\qquad \rank(\Delta\mat{Y})=r\ll n.
\label{eq:low_rank_action}
\end{equation}
where $\mat{U}\in\mathbb{C}^{n\times r}$, $\mat{C}\in\mathbb{C}^{r\times r}$,
and $\mat{V}\in\mathbb{C}^{r\times n}$. A two-terminal branch with no shunt
admittance or line charging has a rank-one outage stamp. In the unit-ratio case,
the stamp is proportional to $\vect{a}\vect{a}^{\T}$, where
$\vect{a}=\vect{e}_m-\vect{e}_n$ and $\vect{e}_k$ is the $k$th canonical basis vector.
A line with nonzero shunt admittance, or a change in transformer tap ratio or phase
shift, produces a general local $2\times2$ update of rank at most two.
Equation~\eqref{eq:low_rank_action} also covers multi-terminal substation reconnections.

Active- or reactive-power set-point changes are a special case of actions. If the generalized admittance matrix is rebuilt,
a change at one bus is a rank-one diagonal update
\begin{equation}
\Delta\mat{Y}=\Delta y_k^{\ext}\vect{e}_k\vect{e}_k^{\T}.
\label{eq:setpoint_diagonal}
\end{equation}
Alternatively, the original generalized matrix can be retained and the set-point deviation represented as an additional nodal correction. The latter avoids an operator update but changes the nonlinear current law. The preferred representation can be selected according to the number and type of actions in a batch.

\subsection{Woodbury Solve}

For either $\mat{A}=\mat{Y}_{\ell\ell}$ or $\mat{A}=\mat{Y}_{qq}$, the update uses the corresponding
restricted action factors. Provided that $\mat{A}$, $\mat{C}$, and the correction
matrix $\mat{K}$ below are nonsingular, the Woodbury identity gives \cite{hager1989updating}
\begin{align}
(\mat{A}+\mat{U}\mat{C}\mat{V})^{-1}
&=\mat{A}^{-1}-\mat{A}^{-1}\mat{U}\mat{K}^{-1}\mat{V}\mat{A}^{-1},\notag\\
\mat{K}&=\mat{C}^{-1}+\mat{V}\mat{A}^{-1}\mat{U}.
\label{eq:woodbury}
\end{align}
Define
\begin{equation}
\mat{Z}=\mathcal S_A(\mat{U}),
\qquad
\mat{W}=\left(\mat{C}^{-1}+\mat{V}\mat{Z}\right)^{-1},
\label{eq:woodbury_precompute}
\end{equation}
where $\mathcal S_A$ is the base-case LU solve. The updated solution for any right-hand side is
\begin{equation}
\vect{x}'=\mathcal S_A(\vect{b})
-\mat{Z}\mat{W}\mat{V}\mathcal S_A(\vect{b}).
\label{eq:woodbury_apply}
\end{equation}
Only the $r\times r$ matrix in~\eqref{eq:woodbury_precompute} is inverted densely,
the post-action admittance matrix itself is not inverted.

When an action touches the slack bus, its non-slack-to-slack entries also
modify the boundary term. Thus,
\begin{equation}
\vect{u}_{\ell}^{0\prime}
=-(\mat{Y}_{\ell\ell}+\Delta\mat{Y}_{\ell\ell})^{-1}
(\mat{Y}_{\ell s}+\Delta\mat{Y}_{\ell s})u_s,
\label{eq:action_zero_voltage}
\end{equation}
where the matrix inversion is handled by applying Equation~\eqref{eq:woodbury_apply} with $\mat{A}=\mat{Y}_{\ell\ell}$.
Similarly, the post-action operator in Equation~\eqref{eq:woodbury_apply} replaces $\mathcal S_{\ell}$ (for $\mat{A}=\mat{Y}_{\ell\ell}$)
and $\mathcal S_q$ (for $\mat{A}=\mat{Y}_{qq}$) in Algorithm~\ref{alg:sparse_pfpd}.

\section{Local-Network Warm Start}
\label{sec:locality}

Localized actions often produce voltage changes that decay with electrical distance. Let $\mathcal K$ be a chosen internal bus set containing every directly affected bus, and let $\mathcal B$ contain all buses outside $\mathcal K$ adjacent to it. The boundary voltages are frozen at their base-case values. The local corrected equation is
\begin{equation}
\mat{Y}_{KK}\vect{u}_K+\mat{Y}_{KB}\vect{u}_B^{\bc}
=\vect{i}_K,
\label{eq:local_kernel}
\end{equation}
with zero-correction voltage
\begin{equation}
\vect{u}_K^0=-\mat{Y}_{KK}^{-1}\mat{Y}_{KB}\vect{u}_B^{\bc}.
\label{eq:local_zero}
\end{equation}
Every boundary bus therefore acts as an additional fixed-voltage slack.
Internal PV and PQ buses use exactly the iteration of Section~\ref{sec:iteration},
with local sparse factors.

Different strategies can be used to defined the internal set $\mathcal K$, for example a country-based definition
in the case of a European Common Grid model. Once this set is identified, exactly the same algorithm~\ref{alg:sparse_pfpd}
can be used with different sets of slack, PV and PQ buses.
The illustrative cases presented in the following section do not rely on local warm starts,
which remains a topic for future exploration.

\section{Experimental Methodology and Results}
\label{sec:results}

\subsection{Test Systems and Benchmark Protocol}

The implementation is assessed on three open access networks: the PEGASE 1354-bus
system, the synthetic Texas grid, and the PEGASE 9241-bus system.
The Texas system is available online and built upon references~\cite{birchfield2017structural,birchfield2018convergence}.
The PEGASE systems are fictitious, structurally representative European transmission cases
\cite{fliscounakis2013contingency,josz2016data}.
\begin{table}[t]
\caption{Networks and N-1 Line-Outage Batches}
\label{tab:systems}
\centering
\footnotesize
\begin{tabular}{lrrr}
\toprule
System & Buses & Line elements & Outages\\
\midrule
PEGASE 1354 & 1354 & 1751 & 134\\
Synthetic Texas & 6717 & 7173 & 200\\
PEGASE 9241 & 9241 & 13797 & 193\\
\bottomrule
\end{tabular}
\end{table}

For each network, a base-case power flow is first solved.
As a deterministic convenience sample, up to the first 200
in-service line elements are then examined, and outages that create an island are
removed before timing. Table~\ref{tab:systems} reports
the pandapower line table used to construct the outage candidates.
pandapower evaluates the retained batch through its
dedicated contingency-analysis module, which dispatches its numba-accelerated
Newton--Raphson implementation over the complete outage batch. PowSyBl is
likewise evaluated through its dedicated AC security-analysis module and the
OpenLoadFlow provider. Thus, neither reference timing is based on an external
Python loop over independent line-outage load flows; each uses the batch
security-analysis facility provided by the corresponding package.

Both timing paths use a power-mismatch tolerance of $0.01$~MVA. In pandapower,
\texttt{tolerance\_mva=0.01} is supplied for the base and N-1 calculations. In the
N-SIDE solver, \texttt{tol\_power=0.01} terminates the iteration when the largest
nodal complex-power gap, after conversion from per unit to MVA, falls below the
same threshold (see Algorithm~\ref{alg:sparse_pfpd}). This establishes a common physical stopping criterion.

The reported values are CPU process times from one batch per system. The base-case
solution is outside the
per-contingency interval for both approaches. For the N-SIDE solver, batch
construction of the outage stamps is included. All runs used pandapower 3.4.0, pypowsybl 1.15.0,
NumPy 2.3.5, and SciPy 1.16.3 on an Apple M4 Pro workstation with 48~GB of unified memory, running macOS Sequoia.

\subsection{Execution Time per Contingency}

Table~\ref{tab:runtime} reports the measured batch total divided by the number of
retained outages. The PowSyBl results use the OpenLoadFlow
provider, with distributed slack and reactive-power limits disabled, and DC voltage
initialization for the base case \cite{powsyblopenloadflow}. The base-case load flow execution time is excluded
from the reported security-analysis time, as for the other two solvers.

\begin{table*}[t]
\caption{Average CPU Time per N-1 Line Contingency}
\label{tab:runtime}
\centering
\footnotesize
\begin{tabular}{lrrrrrr}
\toprule
System & Outages & \shortstack{pandapower\\(ms)} & \shortstack{N-SIDE\\(ms)}
& \shortstack{PowSyBl\\(ms)} & \shortstack{Speedup vs.\\pandapower}
& \shortstack{Speedup vs.\\PowSyBl}\\
\midrule
PEGASE 1354 & 134 & 20.46 & 1.14 & 4.40 & 18.0$\times$ & 3.9$\times$\\
Synthetic Texas & 200 & 66.89 & 4.38 & 39.22 & 15.3$\times$ & 9.0$\times$\\
PEGASE 9241 & 193 & 97.03 & 6.76 & 63.54 & 14.4$\times$ & 9.4$\times$\\
\bottomrule
\end{tabular}
\end{table*}

Relative to pandapower, the N-SIDE solver is faster for all three networks, with
a speedup between $14.4\times$ and $18.0\times$. The similar ratios across
markedly different network sizes indicate that the benefit is not confined to a
single test case.
The absolute N-SIDE time rises from 1.14~ms on PEGASE 1354 to 6.76~ms on
PEGASE 9241, while remaining below 7~ms per outage for every batch. This behavior
is consistent with reuse of the two sparse factorizations: each line outage adds
only a rank-two stamp, a small Woodbury correction, and the nonlinear iterations.

N-SIDE is also faster than PowSyBl on all three networks, from $3.9\times$ on
PEGASE 1354 to $9.0\times$ on Texas and $9.4\times$ on PEGASE 9241. The larger
gains on the two largest systems indicate that reuse of the sparse factors and
low-rank corrections becomes increasingly beneficial as the network solve grows.

The average number of iterations for the N-SIDE solver is 4.12, 3.29, and 5.22 for the
PEGASE 1354, Texas, and PEGASE 9241 grids, respectively.
The time measurements for the different computation phases provide additional insights.
Across the three stored runs, the nonlinear sparse kernel solves account for
approximately 63--75\% of the CPU time, while Woodbury operations account for
approximately 13--26\%. Consequently, further acceleration should primarily
target the repeated sparse triangular solves rather than reconstruction of the
action admittance.

\subsection{Voltage Agreement}

For each retained comparison, the pandapower and N-SIDE solutions are mapped to
the same bus ordering and voltage-angle reference. The recorded metrics are the
maximum, over all buses of the power grid, of the voltage-magnitude difference and of the absolute
voltage-angle difference. The N-SIDE solver uses the 0.01~MVA stopping
criterion described above. For this accuracy-only assessment, pandapower is run
to its tighter default tolerance so that its residual has a negligible effect on
the observed difference. The solved base
case is restored before each pandapower contingency calculation.

\begin{figure*}[t]
\centering
\includegraphics[width=\textwidth]{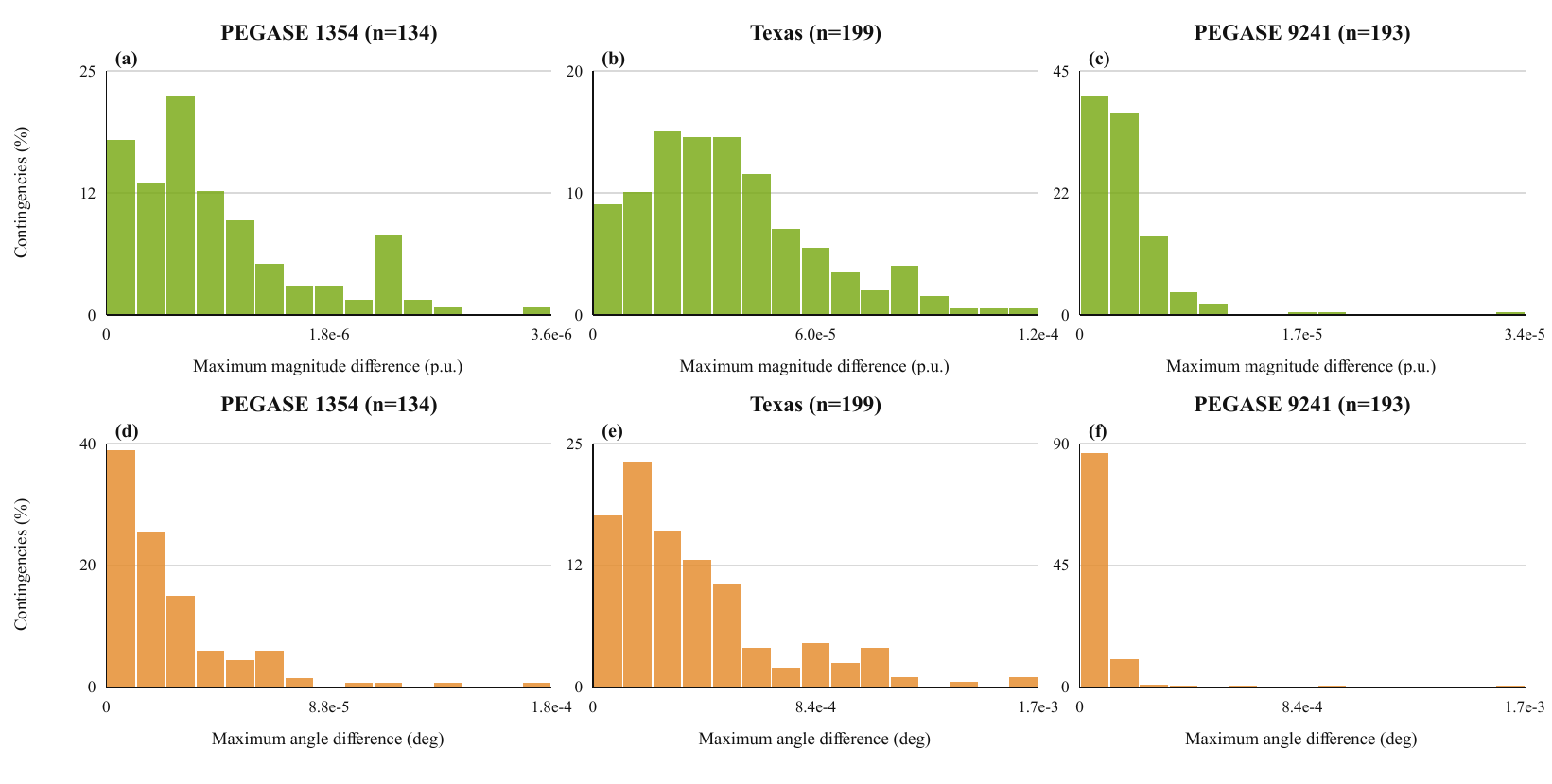}
\caption{Per-contingency maximum bus-voltage discrepancies (the maximum is taken over all buses of the grid) between the N-SIDE
solver and pandapower: (a)--(c) voltage magnitude and (d)--(f) voltage angle for
PEGASE 1354, synthetic Texas, and PEGASE 9241, respectively. Each histogram is
normalized by its number of retained contingencies.}
\label{fig:voltage_discrepancies}
\end{figure*}

Of the 135 non-islanding PEGASE 1354 candidates, the single contingency for
which pandapower and PowSyBl did not converge was removed; all remaining 134
contingencies were retained without further screening. One non-convergent
pandapower case leaves 199 Texas comparisons. For PEGASE 9241, all 193
non-islanding candidates are retained.
Fig.~\ref{fig:voltage_discrepancies} shows the six voltage difference distributions, and Table~\ref{tab:accuracy} gives their principal
quantiles. The N-SIDE solver converges on all retained cases.

\begin{table*}[t]
\caption{Per-Contingency Maximum Voltage Discrepancies}
\label{tab:accuracy}
\centering
\footnotesize
\begin{tabular}{llrrrr}
\toprule
System & Metric & Cases & Median & 95th percentile & Maximum\\
\midrule
PEGASE 1354 & Magnitude (p.u.) & 134 & $6.45\!\times\!10^{-7}$ & $2.28\!\times\!10^{-6}$ & $3.54\!\times\!10^{-6}$\\
PEGASE 1354 & Angle (deg) & 134 & $1.61\!\times\!10^{-5}$ & $6.62\!\times\!10^{-5}$ & $1.72\!\times\!10^{-4}$\\
Synthetic Texas & Magnitude (p.u.) & 199 & $3.27\!\times\!10^{-5}$ & $8.26\!\times\!10^{-5}$ & $1.18\!\times\!10^{-4}$\\
Synthetic Texas & Angle (deg) & 199 & $2.96\!\times\!10^{-4}$ & $1.04\!\times\!10^{-3}$ & $1.65\!\times\!10^{-3}$\\
PEGASE 9241 & Magnitude (p.u.) & 193 & $2.68\!\times\!10^{-6}$ & $8.57\!\times\!10^{-6}$ & $3.35\!\times\!10^{-5}$\\
PEGASE 9241 & Angle (deg) & 193 & $4.87\!\times\!10^{-5}$ & $1.55\!\times\!10^{-4}$ & $1.64\!\times\!10^{-3}$\\
\bottomrule
\end{tabular}
\end{table*}

The two PEGASE distributions are strongly concentrated near zero: their
95th-percentile magnitude differences remain below $9\times10^{-6}$~p.u., and
their median angle differences are below $5\times10^{-5}$ degrees. PEGASE 1354
has maximum magnitude and angle differences of only $3.54\times10^{-6}$~p.u.
and $1.72\times10^{-4}$ degrees, respectively. Texas exhibits broader distributions, but its maximum
magnitude and angle differences remain $1.18\times10^{-4}$~p.u. and
$1.65\times10^{-3}$ degrees, respectively. These differences are small for
security-analysis screening and confirm that the runtime gain is not obtained by
terminating at a visibly different voltage state.

\subsection{Transformer Tap-Position Actions}

\begin{table*}[t]
\caption{Average CPU Time per Five-Step Transformer Tap Action}
\label{tab:tap_runtime}
\centering
\footnotesize
\begin{tabular}{lrrrrrr}
\toprule
System & Actions & \shortstack{pandapower\\(ms)} & \shortstack{N-SIDE\\(ms)}
& \shortstack{PowSyBl\\(ms)} & \shortstack{Speedup vs.\\pandapower}
& \shortstack{Speedup vs.\\PowSyBl}\\
\midrule
PEGASE 1354 & 200 & 7.68 & 2.78 & 19.97 & 2.8$\times$ & 7.2$\times$\\
PEGASE 9241 & 200 & 30.13 & 10.27 & 306.72 & 2.9$\times$ & 29.9$\times$\\
\bottomrule
\end{tabular}
\end{table*}

The broader action formulation is also evaluated for transformer tap changes.
The Texas case is omitted because its source model contains no transformer tap
changer. For each PEGASE network, the first 200 eligible in-service transformers
with a defined, nonzero tap step are selected. Each action increases the initial
tap position by five steps. Because the public PEGASE cases do not specify tap
bounds, these changes constitute a consistent numerical benchmark, while the resulting
tap positions may not be operationally admissible. All 200 actions
converged with each solver on both networks.

Every assessment starts from the solved base-case voltage state. pandapower is
called sequentially with base-case result initialization. PowSyBl is called through
its general AC load-flow API with
previous-value initialization; restoration of the solved base state is performed
before each action and excluded from its action-only timing. The pandapower and
N-SIDE stopping tolerance is again 0.01~MVA, and the remaining solver settings
match the N-1 experiment.

The average number of iterations for the N-SIDE solver is 5.92 and 6.11 for the
PEGASE 1354 and PEGASE 9241 grids, respectively.
Table~\ref{tab:tap_runtime} shows that N-SIDE is approximately three times faster
than pandapower on both PEGASE systems. Its advantage over PowSyBl grows from
$7.2\times$ on PEGASE 1354 to $29.9\times$ on PEGASE 9241. Unlike line outages,
tap-position changes cannot be submitted to the reference solvers through the
dedicated contingency-analysis batch module. The PowSyBl benchmark must
therefore invoke the general-purpose OpenLoadFlow AC solver once per action.
Although the base case solution is given as a warm start, the reconstruction of
the admittance matrix for each action case may explain the substantially longer PowSyBl execution time.
In contrast, N-SIDE prepares
the localized tap admittance changes as one batch, reuses the base sparse factors,
and applies only the corresponding low-rank corrections.

For voltage agreement, pandapower is again run with its tighter default tolerance.
Table~\ref{tab:tap_accuracy} summarizes the voltage discrepancies. All 200 actions are included for
each network. The maximum magnitude discrepancy is $3.90\times10^{-6}$~p.u. on
PEGASE 1354 and $1.06\times10^{-5}$~p.u. on PEGASE 9241; the corresponding
maximum angle discrepancies are $1.83\times10^{-4}$ and
$3.63\times10^{-4}$ degrees.

\begin{table*}[t]
\caption{Per-Action Maximum Voltage Discrepancies for Tap Changes}
\label{tab:tap_accuracy}
\centering
\footnotesize
\begin{tabular}{llrrrr}
\toprule
System & Metric & Cases & Median & 95th percentile & Maximum\\
\midrule
PEGASE 1354 & Magnitude (p.u.) & 200 & $8.68\!\times\!10^{-7}$ & $3.11\!\times\!10^{-6}$ & $3.90\!\times\!10^{-6}$\\
PEGASE 1354 & Angle (deg) & 200 & $2.43\!\times\!10^{-5}$ & $1.06\!\times\!10^{-4}$ & $1.83\!\times\!10^{-4}$\\
PEGASE 9241 & Magnitude (p.u.) & 200 & $2.65\!\times\!10^{-6}$ & $6.31\!\times\!10^{-6}$ & $1.06\!\times\!10^{-5}$\\
PEGASE 9241 & Angle (deg) & 200 & $5.48\!\times\!10^{-5}$ & $2.06\!\times\!10^{-4}$ & $3.63\!\times\!10^{-4}$\\
\bottomrule
\end{tabular}
\end{table*}

\section{Limitations and Conclusions}
\label{sec:conclusion}

This paper reformulates the PFPD corrective-current method so that its principal
linear-algebra operations remain sparse. Treating the slack bus as a fixed-voltage
boundary gives the affine relation in~\eqref{eq:voltage_from_current}; the dense
generator equivalent is replaced by reusable sparse LU factors of
$\mat{Y}_{\ell\ell}$ and $\mat{Y}_{qq}$. Localized admittance changes are then
handled as low-rank updates through the Woodbury identity. On the three study
systems, this combination reduces the measured average N-1 line-contingency time
by factors of 14.4--18.0 relative to pandapower while retaining close agreement
in voltage magnitude and angle. Relative to PowSyBl, it is 3.9--9.4 times faster
across the three systems. For five-step transformer tap actions on the two
PEGASE systems, the measured speedups are 2.8--2.9 relative to pandapower and
7.2--29.9 relative to PowSyBl, with maximum voltage differences below
$1.1\times10^{-5}$~p.u. in magnitude and $3.7\times10^{-4}$ degrees in angle.

As part of the current limitations, reactive-power-limit enforcement is
not supported yet, and all three numerical study cases do not consider the case of
reactive power limits.
The method also has intrinsic numerical limits. Sparse-LU cost depends on fill
and ordering, and Woodbury updates may cease to be advantageous when an action
has high rank or its small correction matrix is ill-conditioned. Islanding
actions  are excluded by the present analysis, as they change the size of the admittance matrix.
Convergence can remain sensitive to the fixed-shunt initialization, as
in the original PFPD method. A robust implementation should consequently retain
a fallback to sparse refactorization and detect ill-conditioned low-rank updates.

Within these limits, the results establish the sparse generalized-admittance
formulation as an effective kernel for high-throughput AC security analysis. Its
fixed circuit representation, reusable sparse factors, and low-rank action
algebra provide substantial speedups without the active/reactive decoupling
assumptions of FDLF. Extending the method to efficiently handle reactive power limits
is the next step toward a broader scope of the proposed load flow solver.

\bibliographystyle{ieeetr}
\bibliography{fast_ac_solver}

\end{document}